\input harvmac
%
\message{S-Tables Macro v1.0, ACS, TAMU (RANHELP@VENUS.TAMU.EDU)}
%
%
\newhelp\stablestylehelp{You must choose a style between 0 and 3.}%
\newhelp\stablelinehelp{You should not use special hrules when stretching
a table.}%
\newhelp\stablesmultiplehelp{You have tried to place an S-Table 
inside another S-Table.  I would recommend not going on.}%
%
%
\newdimen\stablesthinline
\stablesthinline=0.4pt
\newdimen\stablesthickline
\stablesthickline=1pt
%
%
\newif\ifstablesborderthin
\stablesborderthinfalse
\newif\ifstablesinternalthin
\stablesinternalthintrue
\newif\ifstablesomit
\newif\ifstablemode
\newif\ifstablesright
\stablesrightfalse
%
%
\newdimen\stablesbaselineskip
\newdimen\stableslineskip
\newdimen\stableslineskiplimit
%
%
\newcount\stablesmode
\newcount\stableslines
\newcount\stablestemp
\stablestemp=3
\newcount\stablescount
\stablescount=0
\newcount\stableslinet
\stableslinet=0
%
%
%
\newcount\stablestyle
\stablestyle=0
%
%
\def\stablesleft{\quad\hfil}%
\def\stablesright{\hfil\quad}%
%
%
\catcode`\|=\active%
%
%
\newcount\stablestrutsize
\newbox\stablestrutbox
\setbox\stablestrutbox=\hbox{\vrule height10pt depth5pt width0pt}
\def\stablestrut{\relax\ifmmode%
                         \copy\stablestrutbox%
                       \else%
                         \unhcopy\stablestrutbox%
                       \fi}%
%
%
\newdimen\stablesborderwidth
\newdimen\stablesinternalwidth
\newdimen\stablesdummy
\newcount\stablesdummyc
\newif\ifstablesin
\stablesinfalse
%
%
\def\begintable{\stablestart%
  \stablemodetrue%
  \stablesadj%
  \halign%
  \stablesdef}%
\def\stablesadj{%
  \ifcase\stablestyle%
    \hbox to \hsize\bgroup\hss\vbox\bgroup%
  \or%
    \hbox to \hsize\bgroup\vbox\bgroup%
  \or%
    \hbox to \hsize\bgroup\hss\vbox\bgroup%
  \or%
    \hbox\bgroup\vbox\bgroup%
  \else%
    \errhelp=\stablestylehelp%
    \errmessage{Invalid style selected, using default}%
    \hbox to \hsize\bgroup\hss\vbox\bgroup%
  \fi}%
\def\stablesend{\egroup%
  \ifcase\stablestyle%
    \hss\egroup%
  \or%
    \hss\egroup%
  \or%
    \egroup%
  \or%
    \egroup%
  \else%
    \hss\egroup%
  \fi}%
\def\stablestart{%
  \ifstablesin%
    \errhelp=\stablesmultiplehelp%
    \errmessage{An S-Table cannot be placed within an S-Table!}%
  \fi
  \global\stablesintrue%
  \global\advance\stablescount by 1%
  \message{<S-Tables Generating Table \number\stablescount}%
  \begingroup%
  \stablestrutsize=\ht\stablestrutbox%
  \advance\stablestrutsize by \dp\stablestrutbox%
  \ifstablesborderthin%
    \stablesborderwidth=\stablesthinline%
  \else%
    \stablesborderwidth=\stablesthickline%
  \fi%
  \ifstablesinternalthin%
    \stablesinternalwidth=\stablesthinline%
  \else%
    \stablesinternalwidth=\stablesthickline%
  \fi%
  \tabskip=0pt%
  \stablesbaselineskip=\baselineskip%
  \stableslineskip=\lineskip%
  \stableslineskiplimit=\lineskiplimit%
  \offinterlineskip%
  \def\borderrule{\vrule width \stablesborderwidth}%
  \def\internalrule{\vrule width \stablesinternalwidth}%
  \def\thinline{\noalign{\hrule height \stablesthinline}}%
  \def\thickline{\noalign{\hrule height \stablesthickline}}%
  \def\trule{\omit\leaders\hrule height \stablesthinline\hfill}%
  \def\ttrule{\omit\leaders\hrule height \stablesthickline\hfill}%
  \def\tttrule##1{\omit\leaders\hrule height ##1\hfill}%
  \def\stablesel{&\omit\global\stablesmode=0%
    \global\advance\stableslines by 1\borderrule\hfil\cr}%
  \def\el{\stablesel&}%
  \def\elt{\stablesel\thinline&}%
  \def\eltt{\stablesel\thickline&}%
  \def\elttt##1{\stablesel\noalign{\hrule height ##1}&}%
  \def\elspec{&\omit\hfil\borderrule\cr\omit\borderrule&%
              \ifstablemode%
              \else%
                \errhelp=\stablelinehelp%
                \errmessage{Special ruling will not display properly}%
              \fi}%
  \def\stmultispan##1{\mscount=##1 \loop\ifnum\mscount>3 \stspan\repeat}%
  \def\stspan{\span\omit \advance\mscount by -1}%
  \def\multicolumn##1{\omit\multiply\stablestemp by ##1%
     \stmultispan{\stablestemp}%
     \advance\stablesmode by ##1%
     \advance\stablesmode by -1%
     \stablestemp=3}%
  \def\multirow##1{\stablesdummyc=##1\parindent=0pt\setbox0\hbox\bgroup%
    \aftergroup\emultirow\let\temp=}
  \def\emultirow{\setbox1\vbox to\stablesdummyc\stablestrutsize%
    {\hsize\wd0\vfil\box0\vfil}%
    \ht1=\ht\stablestrutbox%
    \dp1=\dp\stablestrutbox%
    \box1}%
  \def\stpar##1{\vtop\bgroup\hsize ##1%
     \baselineskip=\stablesbaselineskip%
     \lineskip=\stableslineskip%
   \lineskiplimit=\stableslineskiplimit\bgroup\aftergroup\estpar\let\temp=}%
  \def\estpar{\vskip 6pt\egroup}%
  \def\stparrow##1##2{\stablesdummy=##2%
     \setbox0=\vtop to ##1\stablestrutsize\bgroup%
     \hsize\stablesdummy%
     \baselineskip=\stablesbaselineskip%
     \lineskip=\stableslineskip%
     \lineskiplimit=\stableslineskiplimit%
     \bgroup\vfil\aftergroup\estparrow%
     \let\temp=}%
  \def\estparrow{\vfil\egroup%
     \ht0=\ht\stablestrutbox%
     \dp0=\dp\stablestrutbox%
     \wd0=\stablesdummy%
     \box0}%
  \def|{\global\advance\stablesmode by 1&&&}%
  \def\|{\global\advance\stablesmode by 1&\omit\vrule width 0pt%
         \hfil&&}%
\def\vt{\global\advance\stablesmode 
by 1&\omit\vrule width \stablesthinline%
          \hfil&&}%
  \def\vtt{\global\advance\stablesmode by 1&\omit\vrule width
\stablesthickline%
          \hfil&&}%
  \def\vttt##1{\global\advance\stablesmode by 1&\omit\vrule width ##1%
          \hfil&&}%
  \def\vtr{\global\advance\stablesmode by 1&\omit\hfil\vrule width%
           \stablesthinline&&}%
  \def\vttr{\global\advance\stablesmode by 1&\omit\hfil\vrule width%
            \stablesthickline&&}%
\def\vtttr##1{\global\advance\stablesmode
 by 1&\omit\hfil\vrule width ##1&&}%
  \stableslines=0%
  \stablesomitfalse}
\def\stablesdef{\bgroup\stablestrut\borderrule##\tabskip=0pt plus 1fil%
  &\stablesleft##\stablesright%
  &##\ifstablesright\hfill\fi\internalrule\ifstablesright\else\hfill\fi%
  \tabskip 0pt&&##\hfil\tabskip=0pt plus 1fil%
  &\stablesleft##\stablesright%
  &##\ifstablesright\hfill\fi\internalrule\ifstablesright\else\hfill\fi%
  \tabskip=0pt\cr%
  \ifstablesborderthin%
    \thinline%
  \else%
    \thickline%
  \fi&%
}%
\def\endtable{\advance\stableslines by 1\advance\stablesmode by 1%
   \message{- Rows: \number\stableslines, Columns:  \number\stablesmode>}%
   \stablesel%
   \ifstablesborderthin%
     \thinline%
   \else%
     \thickline%
   \fi%
   \egroup\stablesend%
\endgroup%
\global\stablesinfalse}
%

\overfullrule=0pt
\abovedisplayskip=12pt plus 3pt minus 3pt
\belowdisplayskip=12pt plus 3pt minus 3pt
%
\def\tilde{\widetilde}

\font\zfont = cmss10 

\def\bigone{\hbox{1\kern -.23em {\rm l}}}
\def\ZZ{\hbox{\zfont Z\kern-.4emZ}}


\lref\senf{A. Sen, {\it ``F-theory and orientifolds''}, 
hep-th/9605150, Nucl.  Phys. {\bf B475} (1996), 562.}
\lref\vafaf{C. Vafa, {\it ``Evidence for F-theory''}, hep-th/9602022,
Nucl. Phys. {\bf B469} (1996), 403.}
\lref\dmm{K. Dasgupta and S. Mukhi, {\it ``Orbifolds of M-theory''},
hep-th/9512196, Nucl. Phys. {\bf B465} (1996), 399.}
\lref\witfive{ E. Witten, {\it ``Five branes and M-theory on an 
orbifold''}, hep-th/9512219, Nucl. Phys. {\bf B463} (1996), 383.}
\lref\bsv{M. Bershadsky, V. Sadov and C. Vafa, {\it ``D-branes and 
topological field theories''}, hep-th/9611222, Nucl. Phys. {\bf B463}
(1996), 420.}
\lref\li{M. Li, {\it ``Boundary states of D-branes and Dy strings''},
hep-th/9510161,  Nucl. Phys. {\bf B460} (1996), 351.}
\lref\djm{K. Dasgupta, D. Jatkar and S. Mukhi, {\it ``Gravitational
couplings and $Z_2$ orientifolds''}, hep-th/9707224.}
\lref\svw{S. Sethi, C. Vafa and E. Witten, {\it ``Constraints 
on low-dimensional string compactifications''}, hep-th/9606122,
Nucl. Phys. {\bf B480} (1996), 213.}
\lref\ghm{M.B. Green, J.A. Harvey and G. Moore, {\it ``I-brane inflow and 
anomalous couplings on D-branes''}, hep-th/9605033,
Class. Quant. Grav. {\bf 14}(1997) 47.}
\lref\sensix{A. Sen, {\it ``Strong coupling dynamics of branes from
M-theory''}, hep-th/9708002.}
\lref\ferrara{S. Ferrara, R. Minasian and  A. Sagnotti, {\it ``Low
energy analysis of M and F theory on Calabi-Yau threefolds''},
 hep-th/9604097,  Nucl. Phys. {\bf B474} (1996), 323.}
\lref\blumzaf{J. Blum and A. Zaffaroni, {\it ``An orientifold from 
F theory''}, hep-th/9607019, Phys. Lett. {\bf B387} (1996), 71.}
\lref\dabpar{A. Dabholkar and J. Park, {\it ``A note on orientifolds and
F-theory''}, hep-th/9607041,  Phys. Lett. {\bf B394} (1997), 302.}
\lref\gp{E. Gimon and J. Polchinski, {\it ``Consistency conditions for
orientifolds and D-manifolds''}, hep-th/9601038, 
Phys. Rev. {\bf D54} (1996), 1667.}
\lref\bersix{M. Berkooz, R.G. Leigh, J. Polchinski, J.H. Schwarz,
N. Seiberg, E. Witten {\it ``Anomalies, dualities and topology of
D=6, N=1 superstring vacua''},  hep-th/9605184, Nucl. Phys. {\bf B475}
(1996), 115.}
\lref\sengp{A. Sen, {\it ``A non-perturbative description of the
Gimon-Polchinski orientifold''},  hep-th/9611186, Nucl. Phys. {\bf B489}
(1997), 139.}
\lref\vafwit{C. Vafa and E. Witten, {\it ``On orbifolds with discrete
torsion''}, hep-th/9409188, J. Geom. Phys. {\bf 15} (1995), 189.}
\lref\gopa{R. Gopakumar and S. Mukhi, {\it ``Orbifold and orientifold
compactifications of F-theory and M-theory to six and four
dimensions''}, hep-th/9607057, Nucl. Phys. {\bf B479} (1996), 260.}
\lref\morsei{D.R. Morrison and N. Seiberg, {\it ``Extremal transitions
and five-dimensional supersymmetric field theories''},
hep-th/9609070, Nucl. Phys. {\bf B483} (1997), 229.}
\lref\abpss{C. Angelantonj, M. Bianchi, G. Pradisi, A. Sagnotti, 
Y.S. Stanev, {\it ``Comments on Gepner models and type I vacua in
string theory''}, hep-th/9607229, Phys. Lett. {\bf B387} (1996), 743.}
\lref\bianchi{M. Bianchi and A. Sagnotti, {\it ``Twist symmetry and open
string Wilson lines''}, Nucl. Phys. {\bf B361} (1991), 519.}
\lref\blum{J.D. Blum, {\it ``F theory orientifolds, M theory
orientifolds, and twisted strings''}, hep-th/9608053,
Nucl. Phys. {\bf B486} (1997), 34.}
\lref\gimjohn{E. Gimon and C.V. Johnson, {\it ``Multiple realisations 
of N=1 vacua in six dimensions''}, hep-th/9606176, Nucl. Phys. {\bf
B479} (1996), 285; {\it ``K3 orientifolds''}, hep-th/9604129,
Nucl. Phys. {\bf B477} (1996) 715.}
\lref\dabpartwo{A. Dabholkar and J. Park, {\it ``An orientifold of 
type-IIB theory on K3''}, hep-th/9602030, Nucl. Phys. {\bf B472}
(1996), 207;
{\it ``Strings on orientifolds''}, hep-th/9604178, Nucl.Phys. 
{\bf B477} (1996), 701.}

{\nopagenumbers
\Title{\vtop{\hbox{hep-th/9709219}
\hbox{TIFR/TH/97-53}}}
{\advance\baselineskip by 6pt
\vtop{\centerline{Anomaly Inflow on Orientifold Planes}}}
\centerline{Keshav Dasgupta\foot{E-mail: keshav@theory.tifr.res.in}
and Sunil Mukhi\foot{E-mail: mukhi@theory.tifr.res.in}}
\vskip 2pt
\centerline{\it Tata Institute of Fundamental Research,}
\centerline{\it Homi Bhabha Rd, Mumbai 400 005, India}
\ \smallskip
\centerline{ABSTRACT}

We examine some six-dimensional orientifold models with $N = 1$
supersymmetry, which can be realised as intersecting 7-branes and
7-planes. These models are studied in the light of recent work showing
that orientifold planes carry anomalous gravitational couplings on
their world-volume. We show that gravitational anomalies can be
locally cancelled by these new couplings at every point in the
internal space, under the assumption that the anomaly residing on
orientifold planes is distributed in a particular way among
brane-plane and plane-plane intersections.

\Date{\vtop{\hbox{September 1997}
\hbox{Revised: March 1998}}}
\vfill\eject} 
\ftno=0
\newsec{Introduction} 

Two types of extended objects have played an important role in string
theory in recent years: Dirichlet branes and orientifold
planes. Classically, there are well-known differences between them: in
contrast to D-branes, orientifold planes are non-dynamical and do not
carry Yang-Mills multiplets on their world volume. However, both kinds
of objects are charged under appropriate $p$-form
potentials. Moreover, in certain models, when quantum corrections are
taken into account, $Z_2$ orientifold planes can split into
nonperturbative generalizations of D-branes\refs{\senf,\morsei}, so
dynamically there is something in common between these two types of
objects.

It has been shown recently that $Z_2$ orientifold planes behave very
much like Dirichlet branes as far as WZ gravitational couplings are
concerned\refs{\djm}. Indeed, both branes and planes carry certain
precise gravitational Wess-Zumino terms on their world-volumes. For
the case of branes, these terms are derived by taking a pair of
intersecting branes and requiring cancellation of gravitational
anomalies on the intersection region via inflow from the
branes\refs{\ghm}. The analogous terms on orientifold planes were
derived in a different way\refs{\djm}, so turning the logic around,
one should be able to check that the predicted WZ terms on planes
actually cancel the anomalies on their intersection regions with
D-branes and with other planes. 

As we will see, it will not be possible to actually demonstrate that
this local anomaly cancellation does take place. Instead, we will
find a prediction for how the WZ terms on intersecting branes and
planes should be distributed among the brane-plane and plane-plane
intersections in order to bring about local anomaly cancellation. An
independent verification of our prediction would be useful in
demonstrating that anomalies really are cancelled locally by the
gravitational couplings found in Ref.\refs{\djm}.

\newsec{Anomalies on Intersections of Branes and Planes}

Let us assume that the gravitational WZ couplings on branes and planes
are of the form
\eqn\wz{\int_B~ C\wedge Y(R)}
where $C$ is the RR background, $Y(R)$ is some curvature polynomial
and $B$ is the world-volume dimension. Generalizing Ref.\refs{\ghm},
we take a configuration of two 7-branes/planes intersecting over
a six dimensional space $B_{12}$. Thus we are considering brane-brane,
brane-plane and plane-plane intersections. The WZ coupling now looks like
\eqn\wess{
-\sum_{i=1}^2 \int_i~(G_1\wedge Y_7 + G_5\wedge Y_3 + Y_0\wedge {}^*
C_0)} 
Here $i=1,2$ labels the world-volume of the two intersecting objects,
each of which can be a 7-brane or a 7-plane. $C_n$ and $Y_n$ are the
background RR $n$-form and curvature $n$-form respectively.  

In the presence of branes or planes, $G_n$ and $dC_{n-1}$ differ. The
former is gauge invariant whereas the latter is not. Hence a partial
integration has been carried out to get Eq.\wess\ from Eq.\wz, using
$Y_8 \equiv dY_7$, $Y_4 \equiv dY_3$.

The relevant Bianchi identities and equations of motion are:
\eqn\bian{
     \eqalign{
dG_1 = \delta^2(1) Y_0(1) + \delta^2(2) Y_0(2)\cr
dG_5 = \delta^2(1) Y_4(1) + \delta^2(2) Y_4(2)\cr
d*G_1 = \delta^2(1) Y_8(1) + \delta^2(2) Y_8(2)}}
{}From the last equation we see that ${}^* C_0$ has an anomalous 
variation
\eqn\anvar{\delta (*C_0) = - \delta^2(1) Y_6(1) - \delta^2(2) Y_6(2)}
where $\delta$ is a general coordinate transformation, and $\delta Y_7
= dY_6$. Therefore, under a general coordinate transformation, the WZ
terms undergo an anomalous variation which can be shown to follow by
the descent procedure from
\eqn\des{-2\int~[Y_8(1)Y_0(2) + Y_8(2)Y_0(1) + Y_4(1)Y_4(2)].}
Thus the anomalies on the intersection regions will be proportional to
the product of the corresponding curvature forms for brane-brane (BB),
brane-plane (BP) and plane-plane (PP) intersections.   

\newsec{Anomaly Cancellation}

The WZ terms on a 7-brane have been determined earlier
in\refs{\ghm,\li,\bsv}. The result is as follows:
\eqn\sevbr{(WZ)_B = \int_{\Sigma^8}~\left[{}^*\tilde \phi -
{1\over 48} C^{(4)+}\wedge p_1 + {1\over 23040} \tilde\phi 
\wedge (9p_1^2 - 8p_2)\right]}
where $\tilde\phi$ and $C^{(4)+}$ are the RR 0-form and 4-form
potentials of the type IIB string, and $p_i$ are the Pontryagin
classes given in terms of the curvature form $R$. The terms on the
orientifold plane have been worked out in \refs{\djm}. In this case
the result is
\eqn\sevpl{(WZ)_P = \int_{\Sigma^8}~\left[-4 {}^*\tilde\phi -
{1\over 24} C^{(4)+}\wedge p_1 + {1\over 11520} 
\tilde\phi \wedge (27p_1^2 - 44p_2)\right]}

The first term in the above equations determine the charges of the
branes and planes. As for the second term in $(WZ)_P$, a different
argument for its coefficient can be given as follows(this is in the
spirit of Ref.\refs{\sensix}). Let us take type IIB on a $T^2/Z_2$
orientifold. Consistency conditions require the existence of 4
orientifold planes and 16 D-branes.  This theory is equivalent to
F-theory on K3\refs{\vafaf,\senf}.

Now, F-theory is conjectured to have a 12-dimensional term of the form
$\int C^{(4)+}\wedge I_8$\refs{\ferrara,\svw}, where $I_8$ is an
8-form polynomial in the curvature. Compactifying this on K3 gives a
term proportional to $\int C^{(4)+}\wedge p_1$ in 8 dimensions. On the
IIB side, let the branes and planes carry the terms $\alpha\int
C^{(4)+}\wedge p_1$ and $\beta\int C^{(4)+}\wedge p_1$
respectively. Since we know that $\alpha= -{1\over 48}$, comparing the
IIB result to that from F-theory gives $\beta = -{1\over 24}$. In
other dimensions, $C^{(4)+}$ is replaced by different RR fields, and
one can show that the WZ couplings (in various dimensions $d = 10-n$)
of branes and planes come with factors $\alpha$ and $\beta$, where
$\alpha$ and $\beta$ are related by $32\alpha + 2^{n+1}\beta =
-1$. The specific case of $d=7$ was treated explicitly in
Ref.\refs{\sensix}.

The third terms in the above equations for the brane and plane are
required to satisfy the conjectured duality of IIB on $T^2/Z_2$ to the
heterotic string on $T^2$\refs{\vafaf,\senf}. This duality implies the
existence of a term $\phi\wedge X_8$ on the heterotic side, where
$X_8$ is another 8-form polynomial in the curvature. ($X_8$ actually
depends on the gauge field strengths as well, but since planes have no
gauge couplings, the gauge fields are set equal to zero for this
discussion.) Since IIB has no such term (its existence would violate
SL(2,Z) invariance), it must come from branes and planes, as has
been shown in Ref.\refs{\djm}.

Before we go on to calculate the inflow contributions $I_{BB}$,
$I_{BP}$ and $I_{PP}$, we should ask what anomalies they are expected
to cancel. What is already known\refs{\ghm} is that the inflow
$I_{BB}$ onto brane-brane intersections cancels the anomalies of the
hypermultiplets which come from Dirichlet-Neumann (DN) open strings
connecting two intersecting branes. In order to investigate anomalies
on BP and PP intersections we need to embed these in a definite
orientifold model, unlike BB intersections, which can be analyzed
independent of a specific model. We examine the Gimon-Polchinski (GP)
orientifold\refs{\gp}\foot{This model and related ones were studied
earlier as open string orbifolds, in Ref.\refs{\bianchi}.} and the
Blum-Zaffaroni-Dabholkar-Park (BZDP) orientifold
model\refs{\blumzaf,\dabpar} to illustrate the cancellation process
explicitly.
\medskip

\noindent{\it GP orientifold}

Since we want a model with intersecting 7-branes and 7-planes, we
consider a T-dual version\refs{\sengp} of the GP orientifold. This is
defined as a $Z_2 \times Z'_2$ orientifold of IIB on $T^4$, where $Z_2
= \{1, g\},~ g = {\cal I}_{67} (-1)^{F_L}\Omega$ and $Z'_2 =
\{1, h\},~ h = {\cal I}_{89} (-1)^{F_L}\Omega$. $T^4$ is a four dimensional
torus labelled by $(x^6,..., x^9)$, $\Omega$ is orientation reversal,
and ${\cal I}_{ij}$ is reversal of the space dimensions $x^i,x^j$.

We now have two sets of orientifold planes, one set at the fixed
points of ${\cal I}_{67}$ and the other set at the fixed points of
${\cal I}_{89}$.  There are four orientifold planes in a set, with
each plane carrying a charge of $-4$ units of the RR scalar
$\tilde\phi$. Cancellation of charges requires placing 16 D 7-branes
transverse to one plane and another set of 16 D 7-branes transverse to
an orthogonal plane. The charge is neutralized locally when each
orientifold plane has four D-branes on top of it. Additionally, $gh =
{\cal I}_{6789} (-1)^F$ will give rise to orbifold twisted sector
states.

We will show that in this model, anomaly inflow onto PP and BP
intersections is necessary to locally cancel the anomalies
which come from the untwisted sector, the orbifold twisted sector and
the brane world-volumes. It will turn out that a specific distribution
of the anomaly on these two types of intersections brings about local
anomaly cancellation; this can perhaps be tested independently in the
future. 

Anomaly cancellation in this model can be viewed in two ways: globally
and locally. Globally, due to overall charge cancellation, there will
be no anomaly inflow and the bulk anomalies will cancel among
themselves. Thus the branes and planes contribute anomaly inflow to
the intersection region in such a way that
\eqn\abc{ n_{BB} I_{BB} + n_{BP} I_{BP} + n_{PP} I_{PP} = 0}
where $n_{BB}, n_{BP}, n_{PP}$ are the number of brane-brane,
brane-plane and plane-plane intersections.

The other aspect, local cancellation, is the emphasis of this
paper. WZ terms on branes are believed to ensure local anomaly
cancellation\refs{\ghm}. In the spirit of the idea that planes (though
not dynamical classically) behave very much like branes, we expect the
analogous result to go through for BP and PP intersections.

A similar situation occurs in the orientifold of M-theory on $T^5/Z_2$
\refs{\dmm,\witfive}. On one hand, due to cancellation of charge, 
there is actually no inflow -- under a general coordinate
transformation the Lagrangian remains invariant. On the other hand, as
observed in Ref.\witfive, the anomalies in the theory are cancelled
locally by inflow from the bulk due to the $C\wedge I_8$ term in the
action.  The presence of five-branes activates the inflow {\it and}
contributes 16 tensor multiplets to the spectrum. This inflow is
reversed by planes carrying {\it minus} half a unit of charge.

One important point about anomaly cancellation in the GP model is that
it is sufficient to cancel the irreducible part of the anomaly, as the
reducible part can be cancelled by an extension of the Green-Schwarz
mechanism\refs{\bersix}. Moreover, for our purposes we can ignore the
observations in Ref.\refs{\bersix} about non-perturbative effects
breaking some $U(1)$ factors, since those issues are irrelevant for
cancellation of the irreducible part of the anomaly. Hence we will
list the perturbative spectrum in what follows.

The spectrum of the T-dual version of the GP model arises as follows.
The untwisted sector consists of one gravity multiplet, one tensor
multiplet and four hypermultiplets of $D = 6, N = 1$
supersymmetry. The twisted sector (coming from both open strings and
orbifold twisted-sector states) consists of vector multiplets and
hypermultiplets. The total spectrum in various regions of the moduli
space is one supergravity multiplet $(g_{\mu\nu}, B^+_{\mu\nu},
\psi_{\mu})$, one tensor multiplet $(B^-_{\mu\nu},
\chi^R, \phi)$, vector multiplets $(A_{\mu}, \psi^L)$ in the adjoint
representation of the enhanced gauge group $G$ at various points in
the moduli space, plus hypermultiplets $(4\phi, \psi^R)$ in various
representations. We list the hypermultiplets along with their origin:

\ \medskip
\begintable
Group $G\times G'$|$U(16)\times U(16)'$|$U(4)^4\times 
U(4)'^4$|$U(2)^8\times U(2)'^8$\eltt
16 fixed points of $gh$|$16\times \bf{(1,1)}$|$16\times
\bf{(1,1)}$|$16\times \bf{(1,1)}$\elt
antisym. rep. of G|$2\times \bf{(120,1)}$|$8\times
\bf{(6,1)}$|$16\times \bf{(1,1)}$\elt
antisym. rep. of G'|$2\times \bf{(1,120)}$|$8\times
\bf{(1,6)}$|$16\times \bf{(1,1)}$\elt
DN open strings|$1\times \bf{(16,16)}$|$16\times \bf{(4,4)}$|$64\times
\bf{(2,2)}$\elt
untwisted sector|$4\times \bf{(1,1)}$|$4\times
\bf{(1,1)}$|$4\times \bf{(1,1)}$
\endtable
\medskip

The DN open string modes are treated separately, as the anomaly from them
is cancelled by inflow from the branes to the intersection
region\refs{\ghm}. Summing over the remaining multiplets at any point
in the moduli space, we find that the irreducible part of the anomaly 
(the coefficient of $\tr R^4$) is equal to ${2\over 45}$. 

Now we can calculate the inflow contribution from the branes and
planes to the intersection region. Combining Eqs.\des,\sevbr\ and
\sevpl, the result for the irreducible part of the anomaly inflow is:
\eqn\inflow{I_{BB} = {1\over 5760},~~ I_{BP} = {7\over 11520},
~~ I_{PP} = -{44\over 5760}.}
Also, it is easy to see that in Eq.\abc, the relevant values are
\eqn\valabc{n_{BB} = 256,~~ n_{BP} = 128,~~ n_{PP} = 16}
satisfying the consistency condition required by charge cancellation.

Using the above results one sees that 
\eqn\canc{
128~I_{BP} + 16~I_{PP} = -{2\over 45}} 
which cancels the anomaly from the spectrum (excluding the DN open
string modes) at all points in the moduli space. Note that the last
term in Eq.\sevpl\ actually does not contribute to this result, hence
global anomaly cancellation does not rely on the existence of that
term. 

As we will see, for anomalies to be cancelled locally, the coefficient
of the last term in Eq.\sevpl\ gets correlated with the proportion in
which bulk anomalies are distributed in the orientifold model. Let us
now examine how the anomalies from the various multiplets are
distributed to the various brane-brane, brane-plane and plane-plane
intersection regions.  We make the following observations:

\noindent(a) For the BB case, as noted above, the inflow cancels the 
anomaly from modes of DN open strings connecting the two branes. In
other words the hypermultiplets in the $(a, b)$ representation of the
group $U(a)\times U(b)'$ lie on this intersection region. Overall at
any point there are 256 hypers, contributing an anomaly of $-{2\over
45}$.

\noindent(b) There are 16 twisted sectors (from the $gh$ part of 
the orientifold group) contributing 16 neutral hypermultiplets. These
are constrained to lie one on each of the 16 plane-plane (PP)
intersection regions. One way to confirm this is to go to the quantum
corrected picture of this model.  As has been shown in
Ref.\refs{\sengp}, the PP intersection region joins smoothly to form a
(nonperturbative) brane. This is due to the presence of blowup modes
of the orbifold fixed points. These twisted sectors contribute a total
anomaly of $16\ldotp -{1\over 5760} = -{1\over 360}$.

\noindent(c) The vectors and the hypers (which come from the 
multiplets on the branes) are confined along the BP intersection
regions. The anomaly from these should be cancelled by inflow from the
intersecting brane and plane. It is easy to check that the difference
between the number of vectors and hypers is 32 at every point in the
moduli space, hence the anomaly from these states is $32\ldotp {1\over
5760}={1\over 180}$.

\noindent(d) At no point in the moduli space can a single brane move 
freely. The minimum number of branes that can move together in this
theory is two, giving the gauge group $U(2)^8\times
U(2)'^8$\refs{\sengp}. In this case the anomaly from BB will be four
times the single BB value, and the anomaly from BP will be double the
original value.

\noindent(e) The only states not accounted for so far are the 
``bulk'' multiplets, from the untwisted sectors, consisting of 1
gravity + 1 tensor + 4 hypers. They contribute a total anomaly of
${1\over 24}$. Because there is no anomaly in the 10d bulk theory, we
must assume that this anomaly, which arises from the orientifolding
operation, is distributed in some way over the orientifold
planes\foot{An analogous assumption was made in Ref.\refs{\witfive},
where the bulk anomaly was distributed equally over 32 orientifold
5-planes.}. This means that it can live on either the BP or the PP
regions. Below we will discover in what proportion it must be
distributed for local anomaly cancellation to take place. 

{}From the above, the total anomalies at the BP and the PP
intersection regions are ${1\over 180}$ and $-{1\over 360}$
respectively. {}From Eqs. \inflow,\valabc\ we know that the inflow
contributions to the BP and PP regions are ${7\over 90}$ and
$-{11\over 90}$ respectively.  Therefore local anomaly cancellation
demands that the untwisted sector multiplets, whose total anomaly is
${1\over 24}$, must have this anomaly distributed in the proportion
$-{1\over 12}$ and ${1\over 8}$ to the BP and PP intersection
regions. Since there are 128 BP and 16 PP intersections, this in
particular implies that each individual BP intersection receives an
anomaly of $-{1\over 1536}$ while each PP intersection receives
${1\over 128}$.

Thus, in the process of arguing that the WZ terms of Ref.\refs{\djm}
are consistent with local anomaly cancellation, we have also made a
prediction: the untwisted sector in the (T-dual) GP orientifold must
have its anomaly localized on BP and PP intersections in the ratio
$-2:3$. An independent confirmation of this prediction would provide
significant support for the idea that anomalies are locally cancelled
in these models.
\bigskip

\noindent{\it BZDP orientifold}

Next we consider a different model in six dimensions with $N= 1$
supersymmetry, which can also be realised in terms of intersecting
7-branes\refs{\blumzaf,\dabpar}. This model has the same orientifold
group $Z_2\times Z_2'$ as the GP model considered earlier, but the
orientation-reversal symmetry $\Omega$ acts with an additional minus
sign on the twisted sector states of the orbifold. (This is like
turning on discrete torsion in the orbifold
construction\refs{\vafwit,\gopa}). This symmetry of the orbifold flips
the sign of the twist fields at all fixed points.

The untwisted sector is the same as before, but now there are no
charged hypermultiplets. They are all projected out. However, the
orbifold twisted sectors contribute 16 tensor multiplets of $N = 1$
supersymmetry. The vector multiplets are in the adjoint of
$SO(8)^4\times SO(8)'^4$.  As this model has no hypers (from the
branes) the moduli of moving the branes are also missing. We now have
2 sets of four intersecting orientifold planes at the fixed points of
the orientifold group. A set of four D-branes lie on top of each
orientifold plane. Therefore we have the following situation. There
are 16 intersection regions. Each region consists of one PP, 16 BB and
8BP intersections. Now one has to calculate the total anomaly from
each such point. The answer is
\eqn\dababc{16 I_{BB} + 8 I_{BP} + I_{PP} = 0}
from Eq.\inflow. Hence in this model there is no net inflow from the
branes and planes. 

The bulk anomalies cancel among themselves, as the spectrum satisfies
the six dimensional anomaly cancellation equation:
\eqn\sixanom{H - V = 273 - 29T}
with $H = 4, V = 224$ and $T = 17$.

Local cancellation in this model is easy to see from the following
observations:

\noindent(1) The 16 intersection regions have a tensor multiplet 
each. These tensor multiplets take the place of hypers in the GP
model, because here the hypers are projected out and tensors are
retained.

\noindent(2) The vector multiplets which are in the adjoint of
$SO(8)^4\times SO(8)'^4$ should lie along the BP intersection
region. But the model has no such {\it distinct} regions. Nevertheless
one can assume the vectors to be distributed equally to the 16
intersection regions of the model, as each one contains a single BP
intersection. 

\noindent(3) For local cancellation of anomalies, an equal fraction of the 
total anomaly from the untwisted sector must go to each of the 16
symmetric intersection regions. Thus we need an anomaly of ${1\over
16}\ldotp {1\over 24}$ at each intersection region. This also follows
from our analysis of the GP model. Recall that there we predicted that
the untwisted sector contributes an anomaly of $-{1\over 1536}$ to
each BP intersection and ${1\over 128}$ to each PP intersection. Since
each intersection region in the BZDP model has 8 BP and 1 PP
intersections overlapping, the anomaly from the untwisted sector will
be $8\ldotp -{1\over 1536} + {1\over 128} = {1\over 384}$ on each such
region, as expected.

\newsec{Conclusions}

The models we have studied here not only exhibit global cancellation
of gravitational anomalies, but are also consistent with local
anomaly cancellation on each of the defect regions. 

We have argued that local anomaly cancellation will take place if
anomalies are distributed in the specific ratio $-2:3$ on brane-plane
and plane-plane intersections. Since we have not found an independent
way of computing this distribution in the present models, our results
do not actually prove that local anomaly cancellation does take place,
but rather should be viewed as a new prediction for the way anomalies
reside on brane-plane and plane-plane intersections.

Although we have only checked two models
explicitly, we expect that all other 6-dimensional orientifold
models\foot{See for example Ref.\refs{\dabpartwo,
\gimjohn, \gopa, \blum, \abpss}.} will exhibit local anomaly 
cancellation in the same way.

Clearly it is important to find an independent way of predicting the
distribution of anomalies onto different types of defect
intersections. This would confirm that the results of Ref.\refs{\djm}
about anomalous couplings on orientifold planes are actually
responsible for local anomaly cancellation. Perhaps more important, it
would give some new insight into the orientifolding procedure itself,
since we do not understand the detailed mechanism by which potential
anomalies are created by orientifolding and eventually cancelled by
inflow from the bulk.
\bigskip

\noindent{\bf Acknowledgements:} We were motivated to carry out this
investigation by suggestions of Dileep Jatkar and Ashoke Sen, which we
gratefully acknowledge. We also thank Atish Dabholkar for helpful
discussions.

\listrefs    

\end